\documentclass[12pt,preprint]{aastex}

\usepackage{CJK}
\begin{document}
\begin{CJK}{UTF8}{gbsn}

\title {A Pilot Search for Evidence of Extrasolar Earth-analog Plate Tectonics}

\author{M. Jura\altaffilmark{a}, B . Klein\altaffilmark{a}, S. Xu\altaffilmark{a}(许\CJKfamily{bsmi}偲\CJKfamily{gbsn}艺) \&  E. D. Young\altaffilmark{b}
}

\altaffiltext{a}{Department of Physics and Astronomy, University of California, Los Angeles CA 90095-1562; jura@astro.ucla.edu, kleinb@astro.ucla.edu, sxu@astro.ucla.edu}
\altaffiltext{b}{Department of Earth Planetary and Space Sciences, University of California, Los Angeles CA 90095; eyoung@ess.ucla.edu}

\begin{abstract}
Relative to calcium, both strontium and barium are  markedly enriched  in Earth's continental crust compared to  the basaltic crusts  of other differentiated rocky bodies within the solar system.     Here, we both re-examine  available archived Keck spectra to place upper bounds on $n$(Ba)/$n$(Ca) and revisit published results for $n$(Sr)/$n$(Ca)  in two white dwarfs that have accreted rocky planetesimals.   We find  that at most only a small fraction of  the pollution is from crustal material that has experienced the distinctive elemental enhancements induced by Earth-analog plate tectonics.  In view of the intense theoretical interest in the physical structure of extrasolar rocky planets, this
search should be extended to additional targets.  
  \end{abstract}
\section{INTRODUCTION}
Internal melting and the subsequent differentiation into an iron-dominated core, a magnesium-elevated mantle and
a silicon-rich crust are central to the evolution of the solar system's rocky planets.  Being on the surface, crusts are the most easily studied of these three different zones, and when examined in detail, they display a wide variety of
elemental compositions  \citep{Taylor2009}.   While other crusts are largely basaltic, Earth's continental crust is compositionally unique  \citep{Rudnick2003} because, as a consequence of  plate tectonics, it is the product of partial melting of preexisting oceanic crustal material -- itself basaltic -- that results in striking enrichments in ``incompatible" elements by magmatic distillation.  

The possibility that extrasolar rocky planets undergo plate tectonics has been extensively discussed \citep{Valencia2007,O'Neill2007, Valencia2009, Kite2009,Korenaga2010,Stamenkovic2011,Stamenkovic2012, vanHeck2011,vanSummeren2011,Foley2012,O'Rourke2012,Stein2013,Bercovici2013}. Much of this effort has been motivated by the recognition that plate tectonics may affect a planet's habitability \citep{Korenaga2012}.   However,
current observational tests  of whether extrasolar rocky planets experience plate tectonics have 
 seemed unrealistic \citep{Lenardic2012}.  Here, we argue that  an indirect  search
for the operation of Earth-analog plate tectonics  on extrasolar rocky planets is possible because of the unique
chemical signature that would be produced in the spectrum of an externally polluted white dwarf.

The generally accepted standard model for the presence of elements heavier than helium in the atmospheres
of white dwarfs cooler than 20,000 K is that these stars have accreted tidally-disrupted asteroids \citep{Debes2002, Jura2003, Jura2014}.  As a zero-order approximation, it has been found that as with bulk Earth,
 major constituents of extrasolar planetesimals are oxygen, magnesium, silicon and iron \citep{Zuckerman2007, Klein2010, Klein2011, Gaensicke2012, Jura2012, Xu2013, Xu2014}. No truly exotic objects as predicted by \cite{Bond2010} -- such as being carbon/silicon dominated or being calcium/aluminum/oxygen dominated -- have been found \citep{Jura2012a}.  Consequently, it
seems that geophysical models for the origin and evolution of rocky objects developed for the solar system may be extended to extrasolar planetesimals.

Evidence is now strong that extrasolar planetesimals have experienced igneous differentiation \citep{Jura2013,Jura2014} with individual
accreted parent bodies understood as measurably consisting of  an Fe-rich core \citep{Gaensicke2012}, an Fe-poor mantle \citep{Farihi2013}, an Al-rich 
crust \citep{Zuckerman2011} or   a mixture of core and crustal material \citep{Xu2013}.  Internal heating and the subsequent melting from the decay of
$^{26}$Al plausibly can help explain these results \citep{Jura2013}. 

 Fragments of evolved planetesimals or even entire planets
may be separated from their parent body by collisions and subsequently accrete onto the white dwarf while retaining
their distinctive compositional signature.  Here, we  assess whether there is even a modest contribution from continental crust-analog
material to the total accretion onto two well observed white dwarfs.
That is, although the bulk of matter accreted onto white dwarfs originates in asteroids, there inevitably must be some collisional erosion of the surface of a rocky planet which would
result in some planetary crustal material also contributing to the pollution of the host star.  In the solar system, there is an exchange among the terrestrial planets of impact ejecta \citep{Gladman1996, Treiman2000}; similar processes must occur in extrasolar environments.

Below, in Section 2, we first describe what we might expect for extrasolar basaltic and continental crustal compositions by extrapolating from the solar system.
Because Sr and Ba over-abundances relative to Ca are uniquely high in Earth's continental crust, we then assess
available evidence for the concentrations of these two elements in two externally-polluted white dwarfs in Section 3.
Finally, in Section 4, we discuss our results and summarize our conclusions.

 \section{CRUSTAL ELEMENTAL COMPOSITIONS: SOLAR SYSTEM}

Although there are many imaginable routes for the formation and evolution of
extrasolar rocky planets,  we develop our expectations of extrasolar crustal compositions
 from those four solar system bodies for which 
detailed abundance measurements of their crusts have been achieved: Earth, the Moon, Mars and Vesta \citep{Taylor2009}.  In all cases,
the two most abundant elements are O and Si, followed by Al and Ca.  Because they are largely sequestered into
the core and mantle, respectively, Fe and Mg are deficient.

Compared to basaltic crusts, Earth's continental crust is compositionally unique:    
``incompatible" elements  are highly
  concentrated within this  outermost zone containing about 0.6\% of the planet's mass  \citep{Rudnick2003}.  Lying in the same column of the periodic table,  Ca, Sr and Ba -- three incompatible elements of interest here -- have relatively large ionic radii  that are accommodated by the more silica and Al-rich tectosilicates that form from crystallization of the partial melts of  oceanic plate material.   The consequent coalescence of the products of melting in the form of island arcs (for example, Japan) produces continental crust that persists because it is too buoyant to be subducted.  The most pronounced chemical enrichment occurs
  for Ba, the largest atom.

 We list in Table 1, the mass concentrations relative to Ca of selected key elements in the four best-studied
 solar system  crusts.    In the crusts of the Moon, Mars and
  Vesta, Sr and Ba are at most modestly enhanced relative to Ca while both elements are markedly enhanced in  Earth's continental crust and modestly depleted in Earth's oceanic crust.  Therefore, simultaneous substantial enhancements of $n$(Sr)/$n$(Ca) and $n$(Ba)/$n$(Ca) can
  serve as an observational signature of Earth-analog plate tectonics.
  \newpage
    \begin{center}
Table 1 -  Crustal  Abundance Ratios by Mass  Relative to Ca
\begin{tabular}{lrrrrrrrr}
\\
Crust  &Mg & Fe &  Sr & Ba & Reference\\
\hline
\hline
Earth - Continental &0.62 & 1.1 & 7.0e-3&0.010& (1)\\
Earth - Oceanic   &   0.62 & 0.89 & 3.0e-4 & 7.3e-5&(2) \\
Lunar Highland &0.24 &  0.31& 1.2e-3 & 6.1e-4& (3) \\
Mars &1.1 & 2.8 & 7.2e-4 &1.1e-3& (3,4) \\
Vesta &0.58 & 2.0 & 1.0e-3&2.3e-4& (5)\\
\hline
CI meteorites &11&20&  8.6e-4 &2.5e-4 & (6)\\
\hline
\end{tabular}
\end{center}
(1) \cite{Rudnick2003}; (2) N-MORB [ = New Mid-Ocean Ridge Basalt] from \cite{Taylor2009}; (3) Ba from \cite{Taylor2009}; (4) Sr from Shergotty meteorite \citep{McSween1985}; (5) Sioux County Eucrite \citep{Mittlefehldt2004} (6) \cite{Wasson1988}

For reference, we also list in Table 1, the abundances in CI chondrites, the assumed approximate initial nebular composition  before the rocky objects formed and evolved. \cite{Reddy2003} reported abundances of 27 elements in a sample of 181 nearby F and G type main sequence stars spanning a wide range of ages.
The Ca, Sr and Ba abundances are almost always within 0.10 dex of the solar value.  Therefore, the approximation that planets form from an
extrasolar nebular with a solar composition is appropriate.  Furthermore, because the condensation temperatures of Ca, Sr and Ba are nearly identical \citep{Lodders2003}, we expect that separation of these elements  occurs in post-nebular physical processing.  The most familiar cosmochemical
pathway for a dramatic enhancement of the strontium and barium abundances occurs during plate tectonics.

\section{STRONTIUM AND BARIUM MEASUREMENTS: EXTRASOLAR PLANETESIMALS}

To date, no purely crustal material has been detected in an externally-polluted white dwarf \citep{Jura2014}.  That is, while
there is good evidence for differentiation, there is no star where Si and O are the dominant pollutants.
Instead, observed accreted material is composed of blends of core, mantle and crust.   Here, we picture
that while most of the accretion onto externally polluted white dwarfs derives from asteroids, there may be collisional erosion of the crust of a planet and this debris could contribute to the total inflow.
\subsection{Model}
 Consider now the standard model to estimate the elemental composition of the parent body (or bodies) from 
measurements of photospheric abundances.  
The mass of the $Z$'th 
element in the outer mixing zone of a white dwarf, $M_{*}(Z)$, is governed by  the rate of mass gain from accretion from a parent body (or parent bodies),   ${\dot M_{PB(S)}}(Z)$, and the rate of mass loss described by settling with the expression \citep{Koester2009}:
\begin{equation}
\frac{dM_{*}(Z)}{dt}\;=\;-\frac{M_{*}(Z)}{t_{Z}}\,+\,{\dot M_{PB(S)}}(Z)\;,
\end{equation}
where  $t_{Z}$ denotes the $Z$'th element's settling time.
We determine spectroscopically $M_{*}(Z)$ in the star's outer mixing zone.

Here, we picture  ${\dot M_{PB(S)}(Z)}$ is dominated by the contribution from asteroids, ${\dot M_{Ast}(Z)}$ but there may be
an additional contribution from a continental crust, ${\dot M_{CC}(Z)}$.  Therefore:
\begin{equation}
{\dot M_{PB(S)}}(Z)\;=\;{\dot M_{Ast}}(Z)\:+\;{\dot M_{CC}}(Z)\;.
\end{equation}

One solution to Equation (1) is that the system is in  a steady state  and the  accretion is ongoing.  In this case, if ${\dot M_{PB(S)}}(Z)$ is constant, the approximate solution  is that:
\begin{equation}
M_{*}(Z)\;{\approx}\;{\dot M_{PB(S)}(Z)}\,t_{Z}\;.
\end{equation}
If so, then for any two elements denoted by $j$ and $k$, the ratio between the masses of detectable material in the parent body (or bodies) is:
\begin{equation}
\frac{{\dot M_{PB(S)}(Z_{j})}}{{\dot M_{PB(S)}(Z_{k})}}\;=\;\frac{M_{*}(Z_{j})}{M_{*}(Z_{k})}\,\frac{t_{Z,k}}{t_{Z,j}}\;.
\end{equation}
In this case, in order to determine the abundances, the relative settling times must be computed in addition to measuring the relative photospheric abundances.

While many different histories can be imagined, the simplest
and most straightforward is that the accretion from both the asteroids and planetary crusts is in a steady state and that Equation (4) pertains.  Consider now a simple two source model where both asteroids with a composition similar to CI chondrites and continental crust material contribute to the accretion.
Define $f_{CC}$, the proportion of continental crust material, as:
\begin{equation}
f_{CC}\;=\;\frac{{\dot M}_{CC}(Ca)}{{\dot M}_{Ast}(Ca)}\;.
\end{equation}
Using the relative abundances in Table 1, we can use Equations (2) and  (5) to write that
\begin{equation}
{\dot M_{PB(S)}}(Ca)\;=\;{\dot M_{Ast}}(Ca)(1\,+\,f_{CC})\;,
\end{equation}
and
\begin{equation}
{\dot M_{PB(S)}}(Ba)\;=\;2.5\,{\times}\,10^{-4}\,{\dot M_{Ast}}(Ca)(1\,+\,40\,f_{CC})\;.
\end{equation}
We combine Equations (6) and (7) to find:

\begin{equation}
\frac{(1\,+\,40\,f_{CC})}{(1\,+\,f_{CC})}\;=\;\left(\frac{{\dot M_{PB(S)}}(Ba)}{{\dot M_{PB(S)}}(Ca)}\right)\left(2.5\,{\times}\,10^{-4}\right)^{-1}\;.
\end{equation}
From measurements of the calcium and barium accretion rates, we  use Equation (8) to solve for $f_{CC}$.
\subsection{Targets}
Here, we consider two of the best studied externally-polluted white dwarfs: GD 362 \citep{Zuckerman2007, Xu2013} and PG 1225$-$079 \citep{Klein2011, Xu2013} where it is already known that the relative abundances are non-chondritic.    Both stars have exceptionally strong Ca II 3933 {\AA} and Ca II 3968 {\AA}  with  a blended  equivalent width larger than 15 {\AA}. 
Because Ca, Sr and Ba all lie in the same column of the periodic table, their ground state resonance lines all lie on approximately the same
curve of growth in the spectra of white dwarfs, and we can measure Sr and Ba abundances
near 10$^{-4}$ that of Ca to make useful comparisons with the ratios listed in Table 1.   On the basis of its having marked excess infrared emission with a prominent 10 ${\mu}$m emission feature, GD 362 is modeled to have a dust disk with an inner temperature greater than 1000 K \citep{Becklin2005,Kilic2005,Jura2007} and therefore likely is mostly accreting from a single asteroid \citep{Jura2008}.   Although PG 1225$-$079 does have a modest infrared excess at 8 ${\mu}$m \citep{Farihi2010}, the inferred maximum dust temperature is only about 300 K.  Likely, there is a large inner hole  in the disk, and   there may be significant contributions to the accretion from more than
one parent body \citep{Farihi2010, Xu2013}.  

We use the same theoretical model atmospheres computed by D. Koester as employed in \cite{Xu2013} and described in \cite{Koester2010}.  We have also adopted the stellar effective temperatures and gravities given in \cite{Xu2013}.  The atomic parameters for Ba are taken from \cite{Mashonkina1999}. The  settling times, as based on the formalism of \cite{Koester2009} and subsequently updated\footnote{time scales can be taken from http://www.astrophysik.uni-kiel.de/~koester/astrophysics/ and further discussed in \cite{Xu2013}}, for Ca, Sr and Ba are computed to equal 9.6 ${\times}$ 10$^{4}$ yr, 4.7 ${\times}$ 10$^{4}$ yr and 3.9 ${\times}$ 10$^{4}$ yr, respectively, for GD 362.  For PG 1225$-$079, the updated settling times
for Ca, Sr and Ba are 1.9 ${\times}$ 10$^{6}$ yr, 
1.4 ${\times}$ 10$^{6}$ yr and 
1.0 ${\times}$ 10$^{6}$ yr, respectively\footnote{There is a typographical error in Table 5 of \cite{Xu2013}; the units for the settling time should have 
been given as 10$^{6}$ yr.}.  The ratios of the settling times used in Equation (4) are more confidently known than their absolute
values \citep{Xu2013} which are not required for this analysis.

At the factor of two level, the chemistry of Ca in extrasolar planetesimals is uncertain \citep{Jura2014}.  Here, we focus on enhancements of between a factor of 10 and 100 which are much greater than the current observational or geochemical uncertainties.
\subsection{Barium Limits}

Barium has not been  detected nor have quantitative upper limits to its abundance previously been reported in any externally-polluted white dwarf.  Therefore, we have re-examined archived Keck/HIRES spectra to place upper limits to the Ba abundance from
 Ba II 4554 {\AA}, as illustrated in Figures 1 and 2.  Using the same model atmosphere analysis as in Xu et al. (2013), we report in Table 2
our derived upper limits both to the Ba abundances  and our estimates of the mass accretion rates of Ba and Ca  from Equation (4).  The previously-measured Ca abundances and inferred results for the relative mass accretion rates of Sr and Ca
 also are provided.

\begin{figure}
 \plotone{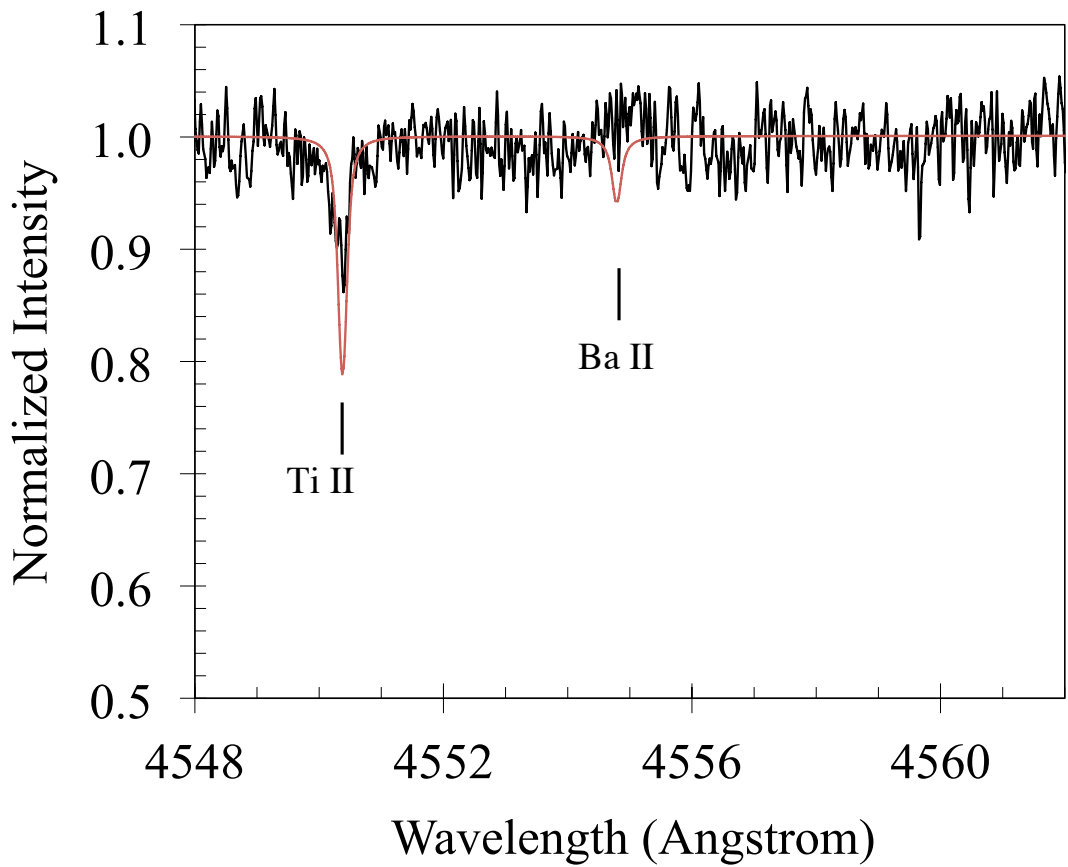}
\caption{Spectrum of GD 362 in the vicinity of Ba II 4554 {\AA}.  Black denotes the data while red denotes the model with our inferred upper limit to the Ba abundance given in Table 2.  The model is wavelength shifted to the photospheric frame  of the star; the equivalent width of Ba II 4554 {\AA} is 18m{\AA}. Wavelengths are in air and the heliocentric frame of rest. } 
\end{figure}

\begin{figure}
 \plotone{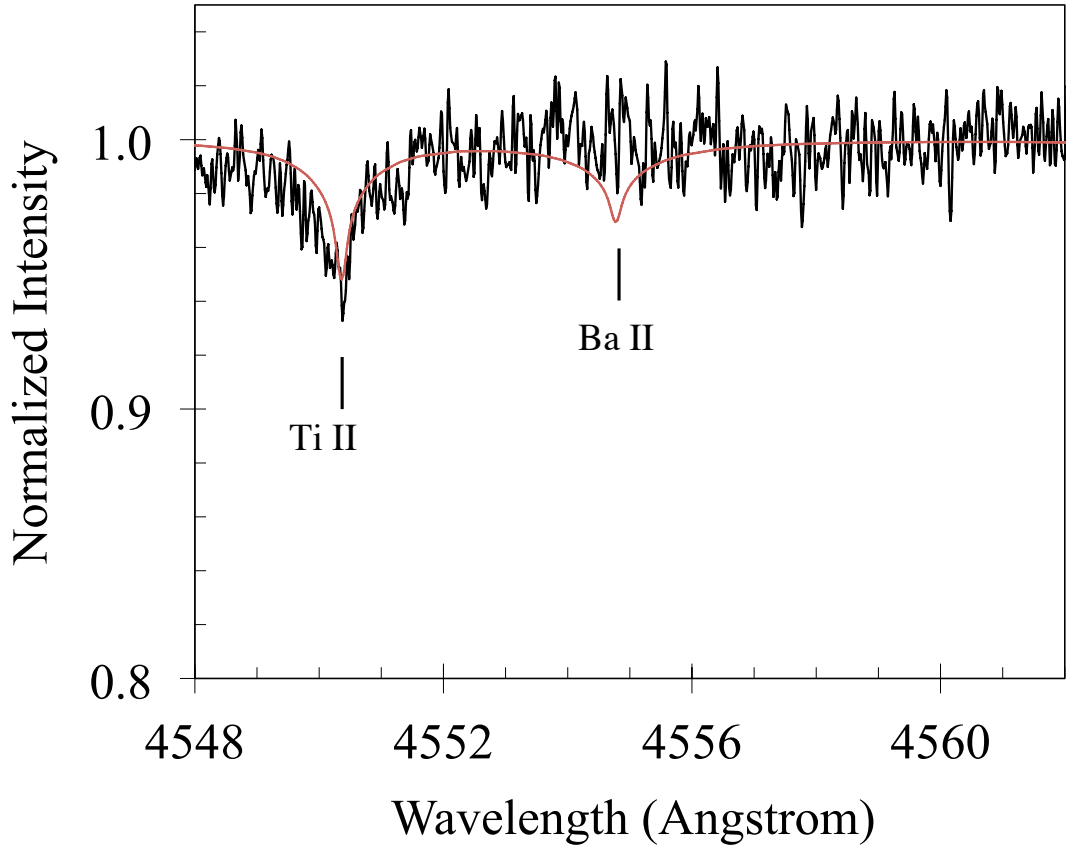}
\caption{The same as Figure 1 except for the spectrum of PG 1225$-$079. Here, the equivalent width of the model Ba II 4554 {\AA} line is 35 m{\AA}.} 
\end{figure}

 \begin{center}
 Table 2 -- Results for Ca, Sr and Ba
 \\
 \begin{tabular}{lrrrrr}
 \\
 Star &  $\log$ $\frac{n(Ba)}{n(He)}$& $\log$ $\frac{n(Ca)}{n(He)}$  & $\frac{{\dot M_{PB(S)}(Sr)}}{{\dot M_{PB(S)}(Ca)}}$ &$\frac{{\dot M_{PB(S)}(Ba)}}{{\dot M_{PB(S)}(Ca)}}$ \\
 \hline
 \hline
GD 362  &  $<$-10.7 & -6.24 & 2.5e-4 & $<$2.9e-4\\
PG 1225$-$079  &  $<$-11.6   & -8.06 & $<$8.8e-4 & $<$1.9e-3\\
 \hline
 \end{tabular}
 \end{center}
We follow standard astronomical usage to report photospheric abundances by number, denoted by $n$.  We follow standard cosmochemical
usage and compare parent body mass ratios, denoted by ${\dot M}$.

The more stringent result is found for GD 362.  For this star, we find from Equation (8) that 
less than 0.4\% of the accreted calcium is from material with the composition of  Earth's continental crust.  Consequently,  the crustal contribution to the contamination of  GD 362  \citep{Xu2013} is mostly basaltic; there is no evidence for any material produced by plate tectonics.  For PG 1225$-$079, the corresponding
result is that less than 20\% of the calcium is derived from material compositionally similar to continental crust.  
\subsection{Strontium Limits}

From Table 2, we find  for GD 362 that the Sr/Ca abundance ratio is even less than in CI chondrites.  This
is not expected in our model and when more data become available, more sophistical models can be developed.  For  PG 1225$-$079,  there is no evidence that   Sr is measurably enhanced  beyond its relative concentration  in CI meteorites.

\section{DISCUSSION}
    At least in  the solar system, the total mass in continental crust is
 greater than the mass in asteroids.  Therefore, even though asteroids dominate the accretion onto white dwarfs,  there might be  favorable circumstances  where the planetary contribution to the heavy elements polluting a white dwarf's photosphere is  measurable.  For example, at Earth,  while most meteorites are derived from asteroids, about 0.07\% originate in the crust of Mars\footnote{http://curator.jsc.nasa.gov/antmet/statistics.cfm: As of 2012, 12 Martian meteorites have been detected out of a collection of 653 achondrites and 17,672 chondrites}.   
 Detailed simulations for a wide variety of  orbital architectures are required to compare the rate of collisional erosion of a planetary crust with the rate of asteroid destruction and, for both processes, 
the subsequent accumulation onto the host white dwarf.   

Consider a strategy to extend the pilot search reported here.
In DZ stars, the equivalent width of Ca II 3933 {\AA} can exceed 30 {\AA} \citep{Sion1990}; these stars typically are identified with low-resolution spectra \citep{Dufour2007, Koester2011}. With follow-up high-spectral resolution studies of  members of this class of stars, it should be possible to make useful measures of the Sr and Ba abundances to compare with different crustal abundances.

\section{CONCLUSIONS}

Available data allow us to conclude, that the abundances of Sr and Ba relative to Ca in two externally-polluted white dwarfs are much
less than found in Earth's continental crust.  Consequently, to-date, there is no evidence for extrasolar Earth-analog plate tectonics.
Given the intense theoretical interest in this topic, additional targets can and should be studied.

This work has been partly supported by the NSF.  We thank D. Koester for computing the model atmospheres and elemental settling times employed here.

 \bibliographystyle{apj}

\end{CJK}
\end{document}